\documentclass[epsf,twocolumn,showpacs,preprintnumbers,superscriptaddress]{revtex4}
\usepackage{graphics}
\usepackage{epsfig}
\begin{document}
\title{Charge Disproportionation and Spin Ordering Tendencies in Na$_x$CoO$_2$ 
} 
\author{K.-W. Lee}  
\affiliation{
Department of Physics, University of California, Davis CA 95616}
\author{J. Kune\v{s}}
\affiliation{
Department of Physics, University of California, Davis CA 95616}
\affiliation{Institute of Physics, Academy of Sciences,
   Cukrovarnick\'a 10, CZ-162 53 Prague, Czech Republic}
\author{W. E. Pickett}
\affiliation{
 Department of Physics, University of California, Davis CA 95616}

\date{\today}
\begin{abstract}
The strength and effect of Coulomb correlations in the 
(superconducting when hydrated) $x\approx$1/3 and ``enhanced'' $x\approx$2/3
regimes of Na$_x$CoO$_2$ are evaluated using the correlated band theory
LDA+U (local density approximation of Hubbard U) method.
Our results, neglecting quantum fluctuations, are:
(1) allowing only ferromagnetic order, there is a critical 
$U_{c}$ = 3 eV, above which charge disproportionation
occurs for both $x$=1/3 and $x$=2/3, 
(2)  allowing antiferromagnetic order at $x$=1/3, $U_{c}$ drops to
1 eV for disproportionation, (3) disproportionation and gap opening
occur simultaneously,
(4) in a Co$^{3+}$-Co$^{4+}$ ordered state, antiferromagnetic
coupling is favored over ferromagnetic, while below $U_{c}$ ferromagnetism
is favored.  Comparison of the calculated Fermi level density of states
compared to reported linear specific heat coefficients indicates 
enhancement of the order of five for $x\sim$0.7, but negligible
enhancement for $x\sim$0.3.  This trend is consistent with strong magnetic 
behavior and local moments (Curie-Weiss susceptibility) for $x>$0.5 while
there no magnetic behavior or local moments reported for $x<$0.5.
We suggest that the phase diagram is characterized by a crossover
from effective single-band
character with $U >> W$ for $x > 0.5$ into a three-band
regime for $x < 0.5$, where $U \rightarrow U_{eff} \leq U/\sqrt{3} \sim W$ 
and correlation effects are substantially reduced.
\end{abstract}
\maketitle

\section{Background}
Since the discovery of high temperature superconductivity in cuprates,
there has been intense interest in transition metal oxides with strongly
layered (quasi) two-dimensional (2D) crystal structures and electronic
properties.  For a few years now alkali-metal intercalated layered cobaltates,
particularly Na$_x$CoO$_2$ (NxCO) with $x\sim 0.50 - 0.75$, 
have been pursued for their thermoelectric properties.\cite{terasaki}
The recent discovery\cite{takada} and 
confirmation\cite{sakurai,euland,lorenz,cao,waki,chou,ornl,wang,schaak}
of superconductivity in the system Na$_x$CoO$_2 \cdot$$y$H$_2$O 
for $x\approx$ 0.3
when intercalated with water at the $y\approx 0.3$ level,
has heightened interest
in the NxCO system.
The structure\cite{ono,lynn,jorgensen} is based
on a 2D CoO$_2$ layer in which edge-sharing CoO$_6$ octahedra lead to a
triangular lattice of Co ions.
Na donates its electron to the CoO$_2$ layer, hence
$x$ controls the doping level of the layer: $x$=0 corresponds to Co$^{4+}$,
S=$\frac{1}{2}$ low spin ions with one minority $t_{2g}$ hole, and $x=1$
corresponds to non-magnetic Co$^{3+}$.  

Nearly all reports of
non-stoichiometric materials quote values
of $x$ in the 0.3 - 0.8 range, and the materials seem generally to show
metallic conductivity.  The $x$=1 endpoint member NaCoO$_2$,
with rhombohedral R$\bar 3$m spacegroup\cite{siegel,takahashi}
is reported to
be a conventional semiconductor.\cite{delmas,shi}  The isovalent compound
LiCoO$_2$ has been more thoroughly studied, with the conclusion that it is
a nonmagnetic band insulator with 2.7 eV bandgap.\cite{elp,czyzyk1} 
The $x=0$ endpoint has been anticipated by many to be a $t_{2g}^5$ Mott
insulator but is less studied; in fact, the Co$^{4+}$ formal oxidation state
in a stoichiometric compound is rare.  The sulfur counterpart CoS$_2$ is
metallic itinerant ferromagnet, close to being half metallic.  
A decade ago, Tarascon and coworkers reported synthesis of 
CoO$_2$ but were unable to identify a complete structure. They concluded  
initially that the Co ions lay on a {\it distorted} triangular 
lattice\cite{amatucci,seguin}.
More recently, further study by
Tarascon {\it et al.}\cite{tarascon} has traced the difficulty in 
pinning down the structure to the existence of two phases of CoO$_2$,
one stoichiometric and the other having 4\% oxygen vacancies.  CoO$_2$
samples are metallic and nonmagnetic, hence cannot be said to contain
Co$^{4+}$ ions.\cite{private}
 
Much has been made of the similarities and differences of this cobaltate 
system compared to the
cuprates.  Both are layered transition metal oxide materials whose
conductivity is strongly anisotropic.  Both are in the vicinity of a
Mott insulator (although the cobaltate one -- CoO$_2$ -- is not well
characterized).  It is possible to vary the carrier concentration ($x$ in
the cobaltate formula) in both systems, with the range in the cobaltates
yet to be agreed on.  In both systems there are specific superconducting
regions: in the cuprates it is a ``dome'' 0.06 $\leq x \leq$ 0.22, 
roughly, while
in the cobaltates there are reports both of a dome 0.27 $\leq x \leq$ 0.33
\cite{schaak} and of a T$_c$=4.5 K plateau for 
0.28 $\leq x \leq$ 0.37.\cite{plateau}
However, the differences between the cobaltates and cuprates
are substantial and expected to be crucial.
Cobalt forms a triangular lattice, which frustrates antiferromagnetic
(AFM) order, while the bipartite square Cu lattice invites AFM order.  
The CoO$_6$
octahedra are edge-sharing, rather than corner-sharing, making the 
bandwidth much narrower and the exchange coupling smaller.  The
cobaltates are electron-doped from the (anticipated) Mott insulator, as 
opposed to the most common hole-doped cuprates.  And most striking,
possibly: in the cobaltates T$_c^{max}$=4.5 K compared to T$_c^{max}$
= 130 K (or higher under pressure) in cuprates.

\begin{figure}[tbp]
\rotatebox{-90}{\resizebox{77mm}{66mm}{\includegraphics{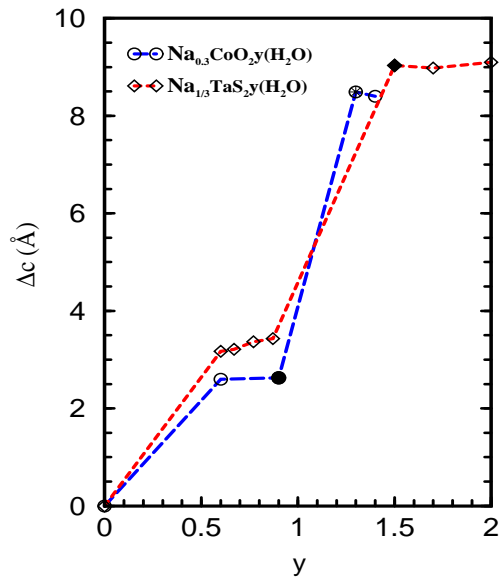}}}
\caption{Change in the $c$ axis lattice parameter with addition of
H$_2$O, in the two systems Na$_{0.3}$CoO$_2$$\cdot$H$_2$O and 
Na$_{1/3}$TaS$_2$$\cdot$H$_2$O, illustrating the great similarity.
For cobaltates, data are from Foo {\it et al.}\cite{foo} (empty circles),
Jin {\it et al.}\cite{ornl} (filled circle), and 
Schaak {\it et al.}\cite{schaak}
(asterisk).  For the chalcogenides, date are from Johnston\cite{DJ2} 
(empty diamonds) and {\it et al.}\cite{DJ3}
(filled diamonds).
}
\label{Delta_c}
\end{figure}

Another system for comparison is the transition metal disulfide based
one, with Na$_{1/3}$TaS$_2$$\cdot y$H$_2$O as the primary comparison.  
In the (Nb,Ta)(S,Se)$_2$ system, 
charge-density waves compete with superconducting
pairing for the Fermi surface, with coexistence occurring in certain cases.
The structure of the (for example) TaS$_2$ layer
is identical to that of the CoO$_2$ layer, consisting of edge-sharing
transition metal chalcogenide octahedrons.  In these 
dichalocogenides, as in the cobaltate system, 
two well defined stages of hydration have been 
identified.\cite{lerf}
In the first stage H$_2$O is incorporated into the same layer as the cation
(typically an alkali ion), and in the second stage two H$_2$O layers are
formed on either side of the cation layer.  The similarity in increase in
the $c$ lattice parameter compared to the cobaltates is illustrated in
Fig. \ref{Delta_c}. 

The electron concentration in Na$_{1/3}$TaS$_2$$\cdot y$H$_2$O
is specified by the Na concentration $x$=$
\frac{1}{3}$, and in this system superconductivity occurs at 4-5 K 
(the same range as in the cobaltates) 
{\it regardless} of the concentration $y$ of water molecules intercalated
into the structure.\cite{DJ1} 
Specific thermodynamically stable phases were identified at
$y$ = 0, $\frac{2}{3}$, 0.8, 1.5, and 2.\cite{DJ2,DJ3,DJ4}
The level of electron donation seems to be
crucial: using Y$_{1/9}$ and La$_{1/9}$ based on the trivalent ions leads
to the same superconducting transition temperature.  Using the divalent
ion Mn ($d^5, S=\frac{5}{2}$), at the same electron donation level
Mn$_{1/6}$TaS$_2$ is {\it ferromagnetic} (FM).
Intercalating this FM compound with H$_2$O leads 
again to T$_c$ $\sim$ 4 K.  This
latter behavior is understood as the water-induced separation of TaS$_2$
layers decreasing the interlayer magnetic coupling sufficiently to 
inhibit long-range magnetic order, thereby allowing
the innate superconducting tendency in the doped TaS$_2$ layer to
assert itself. 

Much of the emphasis, both experimental and theoretical, has been directed
toward the superconducting behavior of the cobaltates, but the long-known
behavior of the tantalum disulfides just mentioned suggests the 
superconductivity may not be so distinctive.
Reports of the magnetic behavior in these cobaltates
have been of particular interest to us.  Except for a charge disproportionated
and charge-ordered phase in a narrow range around $x$=0.5\cite{cavaCO} 
identified by
its insulating behavior, all samples are reported to be metallic.
For $x$ in the 0.5-0.8 range, the susceptibility $\chi(T)$ is
Curie-Weiss-like with reported moment of order
 1 $\mu_B$ per Co$^{4+}$.
\cite{takada,sakurai,ornl}  This local moment is normally interpreted
as indicating the presence of correlated electron behavior on the Co
sublattice, and most theoretical treatments have assumed this
viewpoint.  

Some phase transitions have been reported in the high $x$ region. 
Magnetic ordering at 22 K with small ordered moment
has been reported for $x$=0.75\cite{x75order} based on transport and
thermodynamic data, and the same conclusion was reached by 
Sugiyama {\it et al.} from $\mu$SR studies.\cite{sugiyama1,sugiyama2}
Boothroyd {\it et al.} performed inelastic neutron scattering on $x$=0.75
single crystals and observed FM spin fluctuations.\cite{boothroyd}
Field dependence of the thermopower measured by Wang {\it et al.}\cite{wang} 
indicated that 
the spin entropy of the magnetic Co system ({\it i.e.} the spins of the
Co$^{4+}$ ions) is responsible for the
unusual thermoelectric behavior.  For $x$=0.55, Ando {\it et al.} 
reported\cite{ando}
a rather large linear specific heat coefficient $\gamma$=56 mJ/mol-K$^3$.
Thus for $x > 0.5$ magnetic Co ions
are evident and are strongly influencing the electronic low energy excitations. 

However, for samples with $x \approx $0.3 
({\it i.e.} the superconducting phase) 
many reports concur that the
Curie-Weiss behavior of $\chi$ vanishes.
\cite{sakurai,chou,kobayashi,ornl,comment}.  
In addition, the specific heat $\gamma$ is much smaller, with values
around 12 mJ/mol-K$^2$ reported.\cite{euland,lorenz}
It is
extremely curious that local moments should vanish so near to what 
has been believed 
would correspond to a Mott insulator ($x$=0, Co$^{4+}$ in CoO$_2$),   
and that superconductivity appears only in the moment-free regime.
In the strongly interacting single-band triangular lattice picture, the
$x=0$ system corresponds to the half-filled triangular
lattice that has been studied extensively for local singlet (resonating
valence bond) behavior.\cite{PWA}  The ground state of that model
has however been found to
be N\'eel ordered.\cite{bernu}
In any case, the $x\approx 0.3$ region of 
superconductivity in NxCO is however well away from the expected
Mott-insulating regime,
and the behavior in such systems is expected to 
vary strongly with doping level. 

Much of 
the language used in characterizing this system (above, and elsewhere)
has been based on the 
local orbital, single band picture.
As discussed more fully below, the doping in this system occurs within the
threefold $t_{2g}$ complex of the Co ion, with degeneracy only slightly 
lifted by the non-cubic structure.  The question of single-band versus
multiband nature of this cobaltate system is possibly one of the more
important issues to address, because it can affect strongly the tendency
toward correlated electron behavior. 

Although the primary interest has been in the 
superconductivity of NxCO, there is
first a real need to understand the electronic structure of the normal
state of the unhydrated material, and its dependence on the 
doping level $x$.  The electronic structure
of the $x$=1/2 (with ordered Na) compound in the local 
density approximation (LDA)
has been described by Singh.\cite{singh00,singh03}  Within LDA all Co ions are
identical (``Co$^{3.5+}$''), the Co $t_{2g}$ states are crystal-field split 
(by 2.5 eV) from the $e_g$
states, and the $t_{2g}$ bands are partially filled, 
consequently the system is metallic
consistent with the observed conductivity.  The $t_{2g}$ band complex is
$W \approx$ 1.5 eV wide, and is separated 
from the 5 eV wide O 2$p$ bands that lie just
below the Co $d$ bands.  Singh suggested that the expected on-site Coulomb
repulsion (which has not been calculated) is $U$=5-8 eV on Co,
which gives $U >> W$ so that correlation effects can be
anticipated.

Notwithstanding the experimental evidence for 
nonmagnetic Co ions in the superconducting
material, most of the theoretical 
discussion\cite{ogata1,ogata2,shastry,moess,wangthy,tanaka,honerkamp,bask} has 
focused on the strongly interacting limit, where $U$ is not only important,
but in fact is presumed to be so large that it prohibits 
double occupancy, as described by
the single band $t-J$ model.  The lack of local moments and only weak to
moderate enhancement of the specific heat $\gamma$ 
suggests that a more realistic
picture may be required.  Undoubtedly the single
band scenario is a limited one: although the rhombohedral symmetry 
of the Co site splits
the $t_{2g}$ states into $a_g$ and $e^{\prime}_{g}$ representations, 
the near-octahedral symmetry makes them quasi-degenerate.  Koshibae and
Maekawa have shown that the band dispersion in the 
$t_{2g}$ band complex in these cobaltates 
displays unexpected intricacies, including some analogies to a Kagom\'e
lattice.\cite{koshibae} 

In this paper we begin to address the correlation question by taking
the strongly correlated viewpoint and using
the correlated band theory LDA+Hubbard U (LDA+U) method.  
We investigate two distinct regions
of the phase diagram by focusing on $x =$1/3, the
regime where superconductivity emerges, and $x = $2/3 where more magnetic
behavior is observed.  We find that $U \geq U_c =$ 3 eV 
leads to charge disproportionation (CD) and gap opening for both $x$=1/3 and
$x$=2/3 if only FM order is allowed.  For the N\'eel ordered case at $x$=1/3,
the corresponding transition occurs at $U_c =$ 1 eV.
The availability of three 
distinct sublattices for the ordering, coupled with strong 2D fluctuations, 
may destroy long-range order and make local probes important in studying
charge disproportionation and correlation.
The small values of $U_c$ that we obtain even for $x$=1/3 tend to confuse the
theoretical picture, since there seems to be a conspicuous absence of
correlated electron behavior in this regime of doping.

\section{Structure and Method of Calculation}
\begin{figure}[tbp]
\rotatebox{-90}{\resizebox{8.5cm}{4.7cm}{\includegraphics{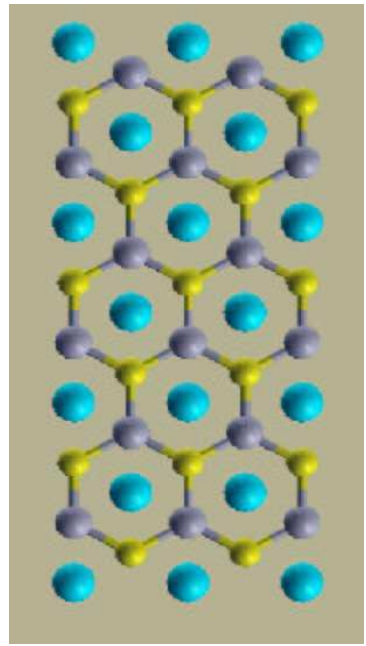}}}
\caption{Illustration of the type of charge disproportionation and spin
ordering that is allowed in the chosen $\sqrt{3}\times\sqrt{3}$
supercell.  The unconnected large spheres represent nonmagnetic Co$^{3+}$
ions, while the large and small connected spheres represent oppositely
directed Co$^{4+}$ spins when ordered antiferromagnetically.
}
\label{AFMsupercell}
\end{figure}

Our calculations are based on the hexagonal structure 
(space group ${P6_{3}22}$),
obtained by Jansen and Hoppe, having lattice constants $(a=2.84 { } \AA, 
c=10.81 { }\AA)$.\cite{Jansen}
The supercell ($\sqrt{3}a \times \sqrt{3}a \times c/2$) is used
so that at the concentration $x=\frac{1}{3}$ that we consider, two 
(or possibly three) inequivalent Co ions, {\it viz.} Co$^{3+}$ and Co$^{4+}$, 
are allowed to emerge in the process of self-consistency.  The allowed
order is displayed in Fig. \ref{AFMsupercell}.
Since we are not analyzing the very small interlayer coupling here, 
a single layer cell is used.
In the supercell (space group $\it{P31m}$, No. 157), atomic coordinations
are Na at the $1a$ (0, 0, 1/2) above/below the Co site at the $1a$ (0, 0, 0),
and the other Co sites are
the $2b$ (1/3, 2/3, 0).  Oxygen sites are the $3c$ (2/3, 0, $\bar{z}_0$)
and the $3c$ (1/3, 0, $z_0$) positions, respectively.
The O height $z_{0} = 0.168 (c/2)= 0.908 { }\AA$, which is relaxed 
by LDA calculation,\cite{singh00}
produces the Co-O-Co bond angle $98.5^\circ $ ($90^\circ $ for undistorted),
so that the octahedra is considerably distorted.

Two all-electron full-potential electronic methods have been used.
The full-potential linearized augmented-plane-waves (FLAPW)
as implemented in Wien2k code \cite{wien2k}
and its LDA+U \cite{nov01,shick} extension were used.
The $s, p$, and $d$ states were treated using
the augmented plane wave+local orbitals (APW+lo) scheme \cite{sjo00}, 
while the standard LAPW
expansion was used for higher $l$'s.
Local orbitals were added to describe Co 3$d$ and O 2$s$ and
2$p$ states.
The basis size
was determined by $R_{mt}K_{max}=7.0$.
The full-potential nonorthogonal local-orbital minimum-basis
scheme (FPLO)\cite{koepernik99,eschrig89}
was also used.
Valence orbitals included Na $2s2p3s3p3d$,
Co $3s3p4s4p3d$, and O $2s2p3s3p3d$.
The Brillouin zone was sampled with regular mesh
containing 50 irreducible k-points.
Both popular forms\cite{ani93,czyzyk} of the LDA+U functional have been used
in our calculations, with no important differences being noticed.

\begin{figure}[tbp]
\rotatebox{-00}{\resizebox{9cm}{9cm}{\includegraphics{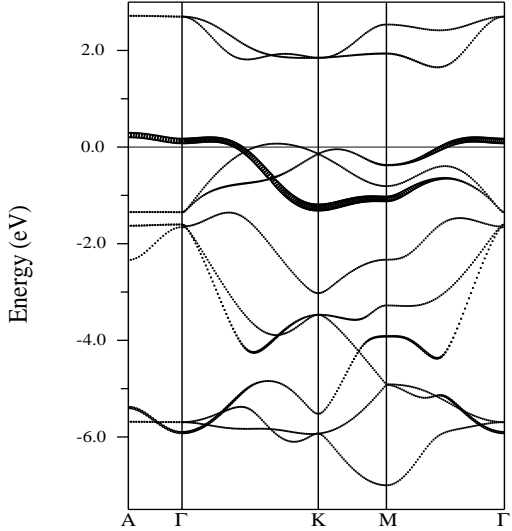}}}
\caption{Local density approximation bands of Na$_{1/3}$CoO$_2$ in the
virtual crystal approximation where there is one Co ion per cell, shown
along the principle high symmetry directions.  The $e_g$ bands lie above
1.5 eV; the bands below -1.5 eV are predominantly oxygen $2p$
in character.  The thickened lines
emphasize the bands within the $t_{2g}$ complex (-1.5 to 0.5 eV)
with strong $a_g$ character.
}
\label{Bands}
\end{figure}

\section{Weakly Correlated Limit.}
{\it LDA electronic structure at $x=\frac{1}{3}$.} 
The $e_g - t_{2g}$ crystal field splitting of 2.5 eV 
that can be seen in the full band
plot in Fig. \ref{Bands} places the (unoccupied) $e_g$ states
(1 eV wide) well out of consideration for most low energy effects.  
The O $2p$ band complex begins just below
the bottom of the $t_{2g}$ bands (see Fig. \ref{Bands}) and is 5.5 eV wide.
The states into which holes are doped from NaCoO$_2$ come from the 1.6 eV wide
$t_{2g}$ band complex.
The trigonal symmetry of the Co site in this triangular CoO$_2$ layer
splits the $t_{2g}$ states into one of $a_g$ symmetry 
[$(|xy>+|yz>+|zx>)/\sqrt 3$ in the local CoO$_6$ octahedron coordinate system]
and a doubly degenerate pair 
$e^{\prime}_{g}$ [$(|xy>+\alpha |yz>+\alpha^2 |zx>)\sqrt 3$ and its
complex conjugate, where $\alpha = exp(2\pi i/3)$].  

\begin{figure}[tbp]
\rotatebox{-90}{\resizebox{6cm}{8cm}{\includegraphics{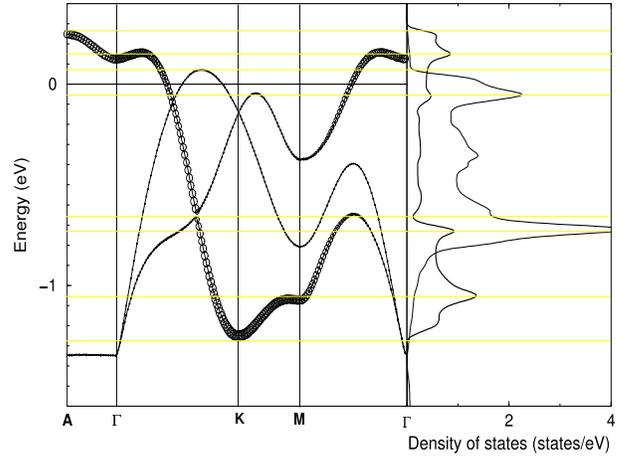}}}
\caption{Band structure (in the virtual crystal approximation)
along high symmetry lines (left panel) and the aligned density of states
(right panel) for the $x=\frac{1}{3}$ cobaltate in the
local density approximation.  The $a_g$ symmetry band is emphasized with
circles proportional to the amount of $a_g$ character.  The $a_g$ density of
states is indicated by the darker line.
}
\label{DOS}
\end{figure}

The $t_{2g}$ band complex that is intersected by the Fermi level E$_F$ 
is shown in more detail in Fig. \ref{DOS},
where the bands with primarily $a_{g}$ character are shown in the ``fatbands''
representation against the corresponding densities of states.  
The band dispersions agree well with those calculated by 
Rosner {\it et al.}\cite{rosner}  The $a_g$ character is strong 
at the bottom of the
$t_{2g}$ complex as well as at the top, and illustrates that holes doped 
into the band-insulating NaCoO$_2$ ($x$=1) phase go initially into one $a_g$
band that is rather flat for $\sim$25-30\% of the distance to the zone boundary.
Based on a rigid band interpretation using this $x$=1/3 density of 
states (DOS), doped
holes enter only $a_g$ states until $x\approx$0.6, whereupon an $e^{\prime}_{g}$
Fermi surface begins to form.  This observation is consistent with the
$x$=0.5 Fermi surface shown by Singh\cite{singh00} 
which has six $e^{\prime}_{g}$
cylinders.

It is of interest to view this band structure from the viewpoint of a
single isolated tight-binding $s$-band on a triangular lattice 
with near neighbor hopping only, which is 
intended\cite{ogata1,ogata2,shastry,moess,wangthy,honerkamp} to model the $a_g$ 
band dispersion.
The $a_g$ DOS lies higher than that of $e^{\prime}_{g}$ not because its 
band center lies 
higher (in fact its centroid is somewhat lower)
but rather due to the particular dispersion and to a substantially larger
effective bandwidth.  Judged from the dispersion curves themselves, the
$a_g$ and $e^{\prime}_{g}$ bands differ little in width.
However, nearly all of the $e^{\prime}_{g}$ states lie within
a 1.0 eV region, whereas the $a_g$ DOS extends over 1.5 eV.

The $a_g$ band dispersion in Fig. \ref{DOS} does resemble that of the simple
tight-binding model $-t\sum_{<ij>}(c_i^{\dag}c_j + h.c.)$ with a negative
value of $t$.  (The band structure also indicates that the $e^{\prime}_{g}$
hopping integral has the opposite sign to that of $a_g$.)
The $a_g$-projected DOS however is nothing like that of the tight-binding
model.\cite{shastry}  The reason is twofold.  
First, there is mixing of the $a_g$ with
the $e^{\prime}_{g}$ bands over most of the Brillouin zone.  The 
hybridization is evident along the M-$\Gamma$ line in Fig. \ref{DOS}; it
is less obvious along the $\Gamma$-K line because the mixing happens to
be accidentally small for the $x$=1/3 CoO$_2$ layer structure.  For other
Co-O distances and bond angles, and for $x$=0 doping level (not
shown), the mixing of the $a_g$
band with the lower $e^{\prime}_{g}$ band becomes much larger.  A second
reason for the actual shape of the DOS is due to the influence of some 
second-neighbor hopping,\cite{rosner} which makes the $a_g$ band near $k$=0 much
flatter than the tight-binding model, or even disperse slightly upward
before turning downward.

Some details of the band structure should be clarified.  The upward dispersion
of the $a_g$ band around the $\Gamma$ point (mentioned above)
also seems to be affected by interlayer
coupling, which can depress the band at k=0.  Johannes and 
Singh have reported that, even for well separated CoO$_2$ layers ({\it i.e.}
when hydrated) the $a_g$ band may still disperse upward\cite{johannes}
before turning down.
Even for CoO$_2$ layer geometries for which there is no upward curvature,
the $a_g$ band 
remains unusually flat out to almost 1/3 of the way to the zone boundary.
Either behavior is indicative of extended hopping processes.

{\it Magnetic Order with LDA}.
Analogous to the results of 
Singh for $x = 0.3, 0.5, 0.7$~\cite{singh00,singh03},
we find ferromagnetic (FM) tendencies for $x$=1/3 within LDA.  
In disagreement with experiment (no magnetic order is observed for 
$x \approx 0.3$, nor even any local moment at high temperature)
a half metallic FM result is found, with a moment of 
2 $\mu_B$/supercell that is
distributed almost evenly on the three Co ions.  The exchange splitting
of the $t_{2g}$ states 
is 0.5 eV, and the Fermi level (E$_F$) lies just above the top of the fully
occupied majority bands (the minority bands are metallic).   
The FM energy gain is about 45 meV/Co.  With the majority bands filled,
the filling of the minority $t_{2g}$ bands becomes $\frac{2}{3}$, leading to
larger $e_g^{\prime}$ hole occupation than for the paramagnetic phase. 
Hence, unlike the standard assumption being made so far, 
$x=\frac{1}{3}$ is a multiband ($a_g$ + $e_{g}^{\prime}$) system (within LDA,
whether ferromagnetic or paramagnetic).
Attempts using LDA to obtain self-consistent charge disproportionated
states, or 
AFM spin ordering, always converged
to the FM or nonmagnetic solution.

{\it Fermiology.}
Suspecting from the $S$=1/2 spins and the two-dimensionality
that quantum fluctuation is an important aspect of
this system, it is possible that the $x$=1/3 system 
is a ``fluctuation-induced paramagnet'' due to the lack of 
account of fluctuations in the electronic structure calculations.
Whatever
the underlying reason, a page can be taken from the high T$_c$ cuprate
chapter of materials physics that, even in the presence 
of considerable correlation
effects, in the magnetically-disordered metallic phase the paramagnetic Fermi
surface (FS) will emerge.  The lack of any observed magnetic behavior in the
$x\approx$0.3 region reinforces this expectation.

\begin{figure}[tbp]
\rotatebox{10}{\resizebox{7.5cm}{7cm}{\includegraphics{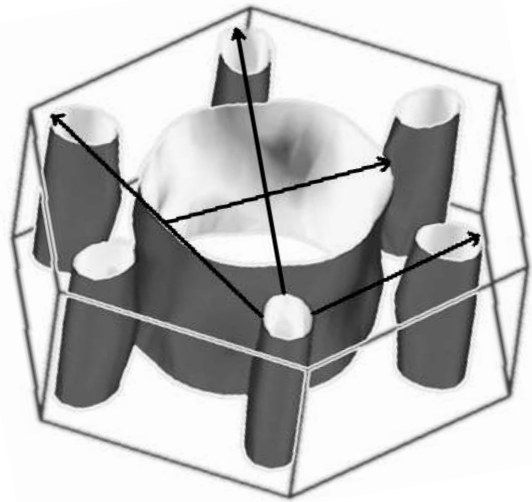}}}
\caption{Fermi surface for (virtual crystal) NxCO, $x$=0.30,
in the two dimensional Brillouin zone.  The large cylinder
contains $a_g$ holes, whereas the six small cylinders contain holes
that are primarily $e_g^{\prime}$-like.
}
\label{FS}
\end{figure}

In Fig. \ref{FS} we show the $x$=1/3 LDA FS,
which is similar to the $x = 0.5$ one shown by
Singh\cite{singh00}.  A large $\Gamma$-centered hole cylinder
(mean radius $K_F$) shows
some flattening perpendicular to the $\Gamma$-K direction, this
cylinder holds 0.43 $a_g$ holes/Co.  In addition, there are six
additional, primarily $e_g^{\prime}$ in character, hole cylinders
lying along the $\Gamma$-K directions, containing 0.04  holes in each of the
six small cylinders (radius $k_F$).  The total is the
0.67 holes necessary to account for the $x=0.33$ electron count.
This FS geometry leads to several important
phase space features.  There are the nesting wavevectors that translate
one of the small cylinders into another, giving three distinct
intercylinder nesting vectors as illustrated in Fig. \ref{FS}.  
If these cylinders were circular, these
vectors would represent strong nesting vectors for charge- or spin-density
waves.  In addition, the susceptibility for $Q \leq 2k_F$ 
intra surface scattering
processes is constant in two dimensions.\cite{2Dphasespace}
The calculated cylinders have
an eccentricity of 1.25, weakening these nesting features somewhat.
There are in addition the corresponding
processes with $Q \leq 2K_F$ of the large
cylinder.

\begin{figure}[tbp]
\rotatebox{-90}{\resizebox{7cm}{6cm}{\includegraphics{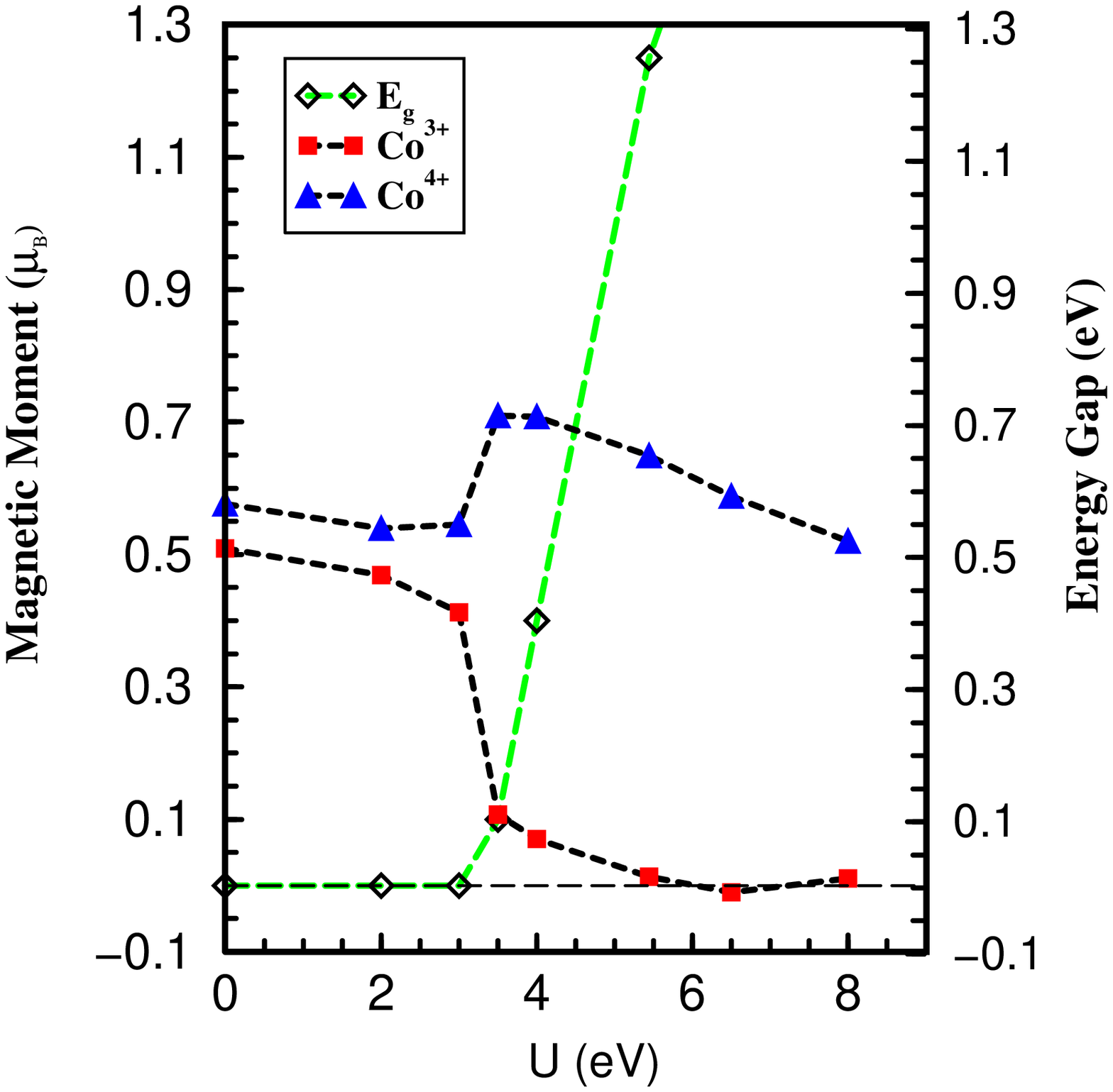}}}
\rotatebox{-90}{\resizebox{7cm}{6cm}{\includegraphics{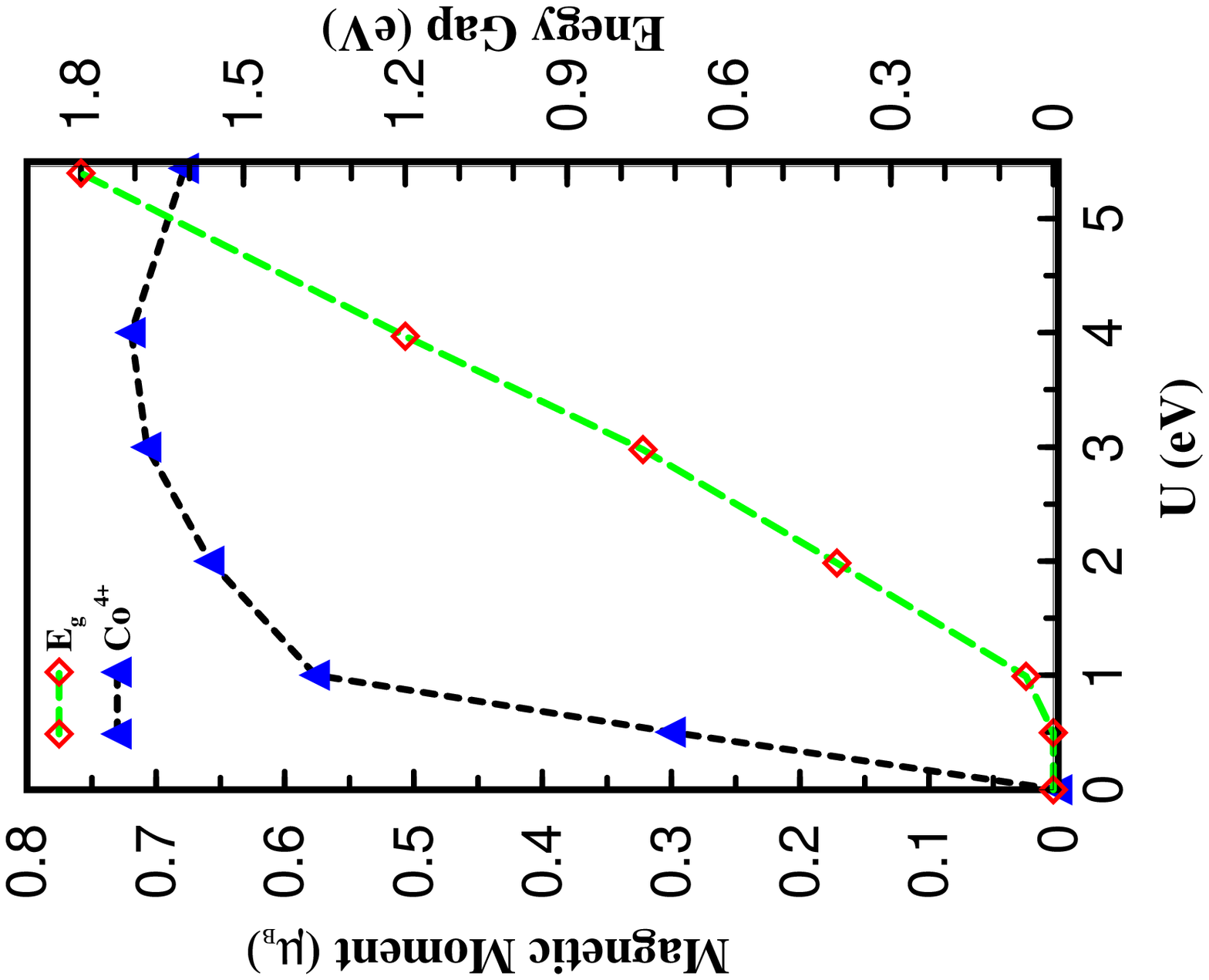}}}
\caption{Effect of the intraatomic repulsion $U$ on the magnetic 
moment (left axis) and energy gap (right axis) of the Co1 and Co2 ions 
in the supercell.  Top panel: result for ferromagnetic order.  
Changes in the magnetic moments, indicating disproportionation to formal
charge states Co$^{3+}$ and Co$^{4+}$, begins at $U_c$ = 3 eV.
The opening of the gap (half metallic to insulating) occurs simultaneously
at $U_{c}$. Bottom panel: results for antiferromagnetic order.  Already
for U$_{c}$=1 eV, the Co$^{4+}$ moment is large (the Co$^{3+}$ moments is
zero by symmetry) whereas the gap is just 
beginning
to open.
}
\label{Moments}
\end{figure}

\section{Inclusion of Correlation Effects}
Despite the feature of the LDA+U method that it drives local orbital
occupations to integral occupancy (as $U$ increases), to our knowledge it
has never been used to study the question of charge disproportionation.
In this section we show that moderate values of $U$ lead to CD at both
$x$=1/3 and $x$=2/3. For the triangular lattice, threefold expanded 
supercells ($\sqrt 3 \times \sqrt 3$, see Sec. II) 
are convenient, and $x$=1/3 lies very
close to the superconducting composition while $x$=2/3 is representative of
the $x >$ 0.5 region that shows correlated behavior.\cite{zou_ldau}  

{\it LDA+U magnetic structure and energies.}  The behavior of the 
LDA+U results (Co moments and the energy gap)
versus $U$ was first studied for $x$=1/3 
(on-site exchange was kept fixed 
at the conventional value of 1 eV as $U$ was varied).  The dependence of the 
magnetic moment and band gap on $U$ 
for FM ordering is shown in the top panel of Fig. \ref{Moments}.  
For $U < U_c$ = 3 eV, the moments on the two inequivalent Co sites
are nearly equal and similar to   
LDA values (which is the $U\rightarrow$0 limit).  Above $U_c$,  
disproportionation into $S = \frac{1}{2}$ Co$^{4+}$ and
$S = 0$ Co$^{3+}$ ions is nearly complete at $U$ = 3.5 eV and is
accompanied by a metal-insulator (Mott-like) transition from conducting to
insulating.  The gap increases linearly at the rate d$E_g$/d$U$ = 0.6 for
$U >$ 3.5 eV.
For the (insulating) $U$=5 eV case, nonmagnetic Co$^{3+}$ states 
lie at the bottom of the
1.3 eV wide gap, with the occupied Co$^{4+}$ 
$e_{g}^{\prime}$ states 1-2 eV lower. 
The spin-half ``hole'' on the Co$^{4+}$ ion occupies the $a_g$ 
orbital as expected.  

In our choice of (small) supercell, 
this CD is necessarily accompanied by
charge order, resulting in a honeycomb lattice of spin half ions.
In a crystal there would be three distinct choices of the ordered sublattice
(corresponding to the three possible sites for Co$^{3+}$). 
Of course, even at a rational concentration such as $x$=1/3, CD
may occur without necessarily being accompanied by 
charge ordering when thermal and
quantum fluctuations are accounted for.  Regarding the disproportionation,
we note that, based on
the Mullikan charge decomposition in the FPLO method, the charges on
the ``Co$^{3+}$'' and ``Co$^{4+}$'' ions differ by only 0.25-0.3 electrons.
This small value reflects the well known result that the formal charge
designation, while being very informative of the magnetic state and
indispensable for physical understanding, does not represent actual
charge accurately.

\begin{figure}[tbp]
\rotatebox{-90}{\resizebox{7cm}{6cm}{\includegraphics{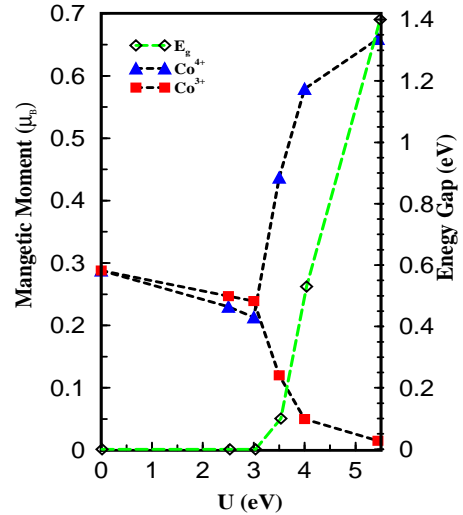}}}
\caption{Effect of the intraatomic repulsion $U$ on the magnetic
moment (left axis) and energy gap (right axis) of the Co1 and Co2 ions
in the supercell for $x$=2/3.  
Changes in the magnetic moments, indicating disproportionation to formal
charge states Co$^{3+}$ and Co$^{4+}$, begins at $U_c$ = 3 eV just as
in the corresponding $x$=1/3 case.  Note that applying small but increasing
$U$ decreases the moment somewhat until disproportionation occurs. 
The opening of the gap (half metallic to insulating) occurs simultaneously
at $U_{c}$. 
}
\label{Moments2}
\end{figure}

The analogous calculation can be carried out allowing AFM order of the
Co$^{4+}$ ions, and the results are shown in the bottom panel of Fig.
\ref{Moments}.  A Co$^{4+}$ moment grows (disproportionation)
immediately as $U$ is increased
from zero.  The gap opens around $U$ = 1 eV and increases at the rate
d$E_g$/d$U$ = 0.4.  Thus for AFM spin order, the critical value is no
higher than $U_c$ = 1 eV.
 
At $x$=2/3, CD will lead to only one Co$^{4+}$ ion in the supercell, so
only FM ordering can be considered.  The corresponding behavior of the moment 
and the gap are presented in Fig. \ref{Moments2}.  The Co moments remain
nearly equal but slowly decrease from their LDA value up to $U_c$
= 3 eV, whereupon again CD occurs abruptly.  The moments are ``well
formed'' by $U$ = 4 eV but continue to evolve somewhat beyond that.
In this case d$E_g$/d$U$ = 0.67.

\begin{figure}[tbp]
\rotatebox{-90}{\resizebox{8cm}{8cm}{\includegraphics{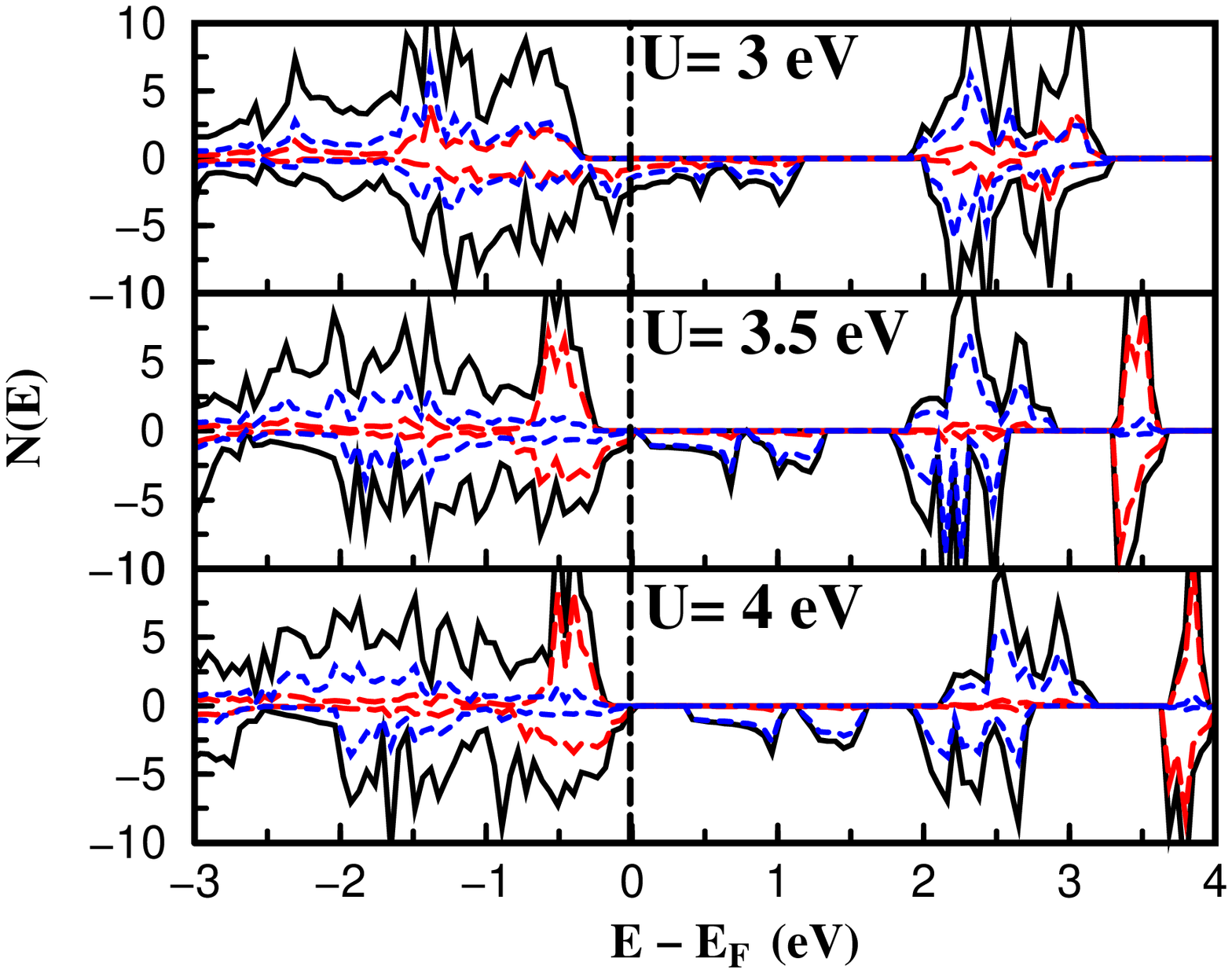}}}
\rotatebox{-90}{\resizebox{8cm}{8cm}{\includegraphics{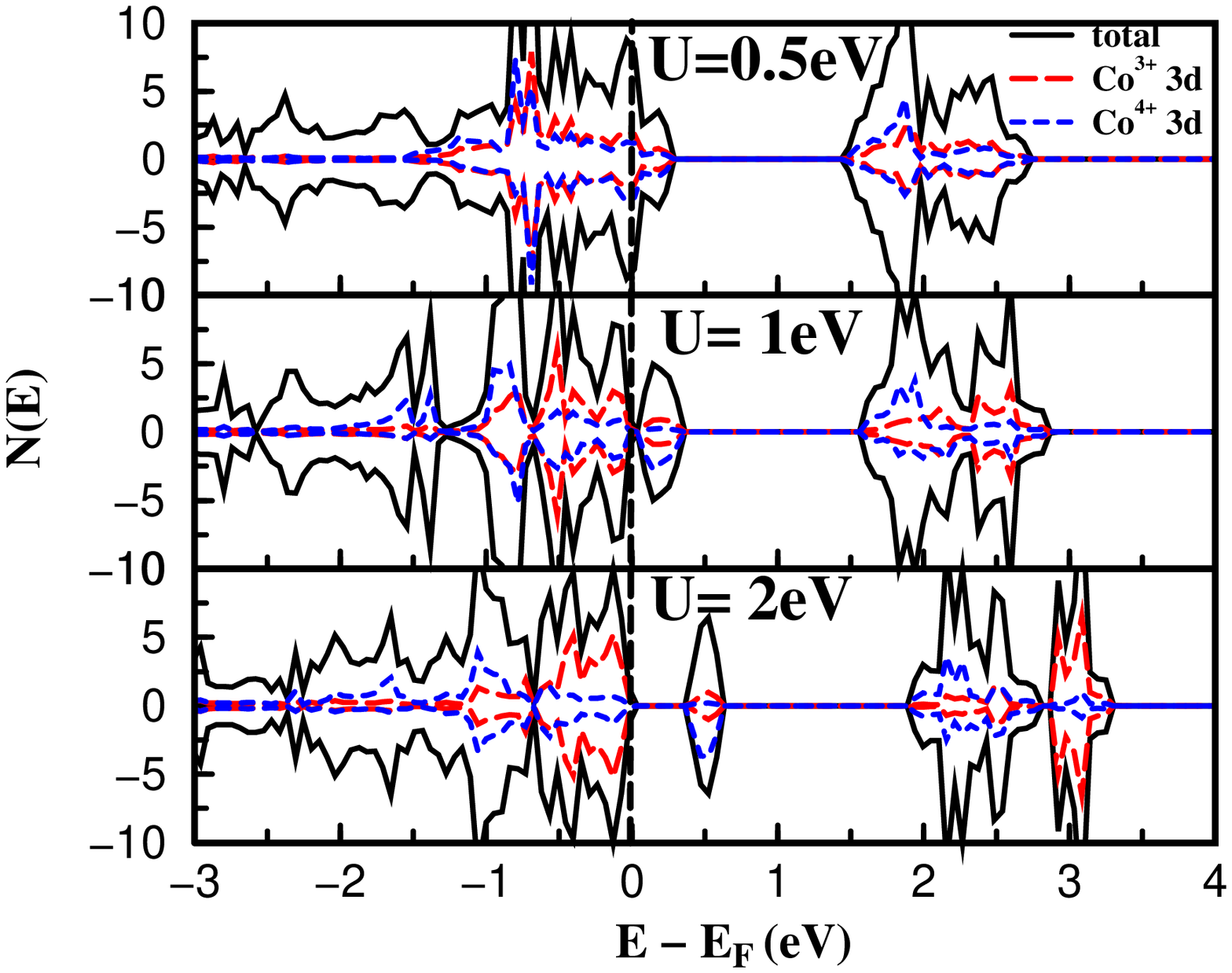}}}
\caption{Evolution of the Co$^{3+}$ and Co$^{4+}$ $3d$ states as charge
disproportionation occurs, for $x$=1/3 and FM spin order (top panel) and
AFM spin order (bottom panel).  Note that the evolution in the FM case is
from a half metallic FM.
}
\label{dDOS_vs_U}
\end{figure}

Our LDA+U results, showing charge disproportionation for $U_c$ = 3 eV (FM)
or $U_c$ = 1 eV (AFM) are very different from earlier reports where
little change was obtained even for larger values of U when symmetry
breaking by disproportionation was not allowed.\cite{zou_ldau}  This
difference serves as an alert that LDA+U results can be sensitive to what
degrees of freedom are allowed.

It is far from obvious that charge disproportionation and gap opening should
occur simultaneously with LDA+U, although physically the phenomena are
expected to be closely related.  
The evolution of the Co$^{3+}$ and Co$^{4+}$ $3d$ states with $U$ in the
critical region is shown in Fig. \ref{dDOS_vs_U}.  In the FM case,  the
system evolves from a half metallic configuration, still visible for $U$
= 3 eV, and gap opening occurs just
when the minority $a_g$ band containing 2/3 hole per Co
ion splits off from the valence
band.  At the point of separation, the $a_g$ states split into an unoccupied
narrow band containing one hole for each of the two Co$^{4+}$ ions, and an
occupied (and also narrow) band on the Co$^{3+}$ ion.  This disproportionation
can be identified from the strong change in the {\it occupied} states around
-0.6 to -0.3 eV.  Thus while disproportionation in principle
could occur before the
gap opens, it does not do so here.
 
When symmetry is broken according to the AFM ordered (and disproportionated)
superlattice shown in Fig. \ref{AFMsupercell}, the critical value of $U$ is
1 eV or less.  This ``ease'' in gap opening is no doubt encouraged by the
narrow bandwidth of the unoccupied $a_g$ band.  A spin up Co$^{4+}$ ion is
surrounded by three spin down Co$^{4+}$ ions and three Co$^{3+}$ ions, 
neither of which have $a_g$ states of the same spin direction at the same
energy.  The surviving bandwidth reflects the effective coupling between
$a_g$ states on second-neighbor Co ions.  For this AFM ordered phase,
the DOS of Fig. \ref{dDOS_vs_U} indicates also little or no
disproportionation before the gap begins to open.

{\it Exchange Coupling.}
The Coulomb repulsion parameter $U$ has not been calculated for these
cobaltates, but
several estimates for other cobaltates put $U$ at 5 eV or 
above.\cite{singh00,ogata2,tanaka}  On the other hand, Chainani {\it et al.} fit
cluster calculation results to xray photoemission data on samples showing
both charge states, and found that $U$ in the 3-5.5 eV range work
equally well.\cite{chainani}  This range is rather lower than what has
generally supposed, thus the appropriate value of 
$U$ is quite uncertain.
Here we concentrate on  
$U$=5.4~eV results, but our calculated behavior is not sensitive to 
variation of $U$ in this range. 
In this CD regime (also charge ordered, due to the constraint of the 
supercell), 
AFM ordering gives 
1.2 mRy/Co lower energy than does FM order.
In terms of nearest neighbor coupling on the resulting honeycomb lattice, 
the FM - AFM energy difference corresponds
to $J$ = 11 meV.  Referring to the paramagnetic bandwidth identified 
above, the corresponding superexchange constant would be 4$t^2/U$
$\sim 20$ meV.  Again, we note that the $a_g$ DOS differs greatly from
the single band picture that was used to obtain this value of 
$t$=0.16 eV.

\section{Discussion of Interaction Strength} 
Our calculations indicate that, as long as $U \geq$ 3 eV as is generally
thought, at both $x$=1/3 and $x$=2/3 there is a 
strong tendency to disproportionate, with one result being a Co$^{4+}$ ion
with a local moment.  Disproportionation into an AFM honeycomb lattice
occurs already by $U_c$ = 1 eV in our mean field treatment. 
At least in the presence of
charge order in a honeycomb arrangement, there is an AFM nearest neighbor
exchange coupling $J \approx 11$ meV. 
The N\'eel state is known to be the ground state
of the AFM Heisenberg honeycomb lattice. 
These charge ordering tendencies may be expected to be 
strongly opposed by thermal and quantum
fluctuations that are expected for low coordination and small spins in
2D layers.  To date, disproportionation and charge ordering have only been
reported for $x$=1/2.\cite{cavaCO} 

In addition to producing local magnetic moments, charge disproportionation
might be expected to introduce new 
coupling to the lattice.  Since the radii of Co$^{2+}$ and Co$^{3+}$
differ by 15\% (0.74 \AA~ vs. 0.63 \AA), disproportionation into those
charge states would be expected to couple strongly to local oxygen modes.  
In octahedral coordination, however, the Co$^{4+}$ ion radius is almost
indistinguishable from that of
the Co$^{3+}$ ion,\cite{ionicradii} so there may be little evidence in the
lattice behavior even if Co$^{3+}$-Co$^{4+}$
charge disproportionation occurs.

In spite of the prevalent theoretical presumption that correlation effects 
may be playing an essential role in the superconductivity of this
cobaltate system, the data seem to be suggesting otherwise.  In the $x >$
0.5 regime, indeed local moments are evident in 
thermodynamic and transport data,
spin fluctuations have been observed by neutron scattering, and the linear
specific heat coefficients are large, $\gamma$=48-56 mJ/mol-K$^2$.  
Comparing this value to our
calculated band value, $\gamma_{\circ}$ = 10$\pm$2 
mJ/mol-K$^2$, leads to a factor of five enhancement due to dynamic 
correlation effects.  Magnetic ordering around $x\sim 0.75$ also attests
to substantial correlation effects. 
Our finding of disproportionation for $U >$ 3 eV (in mean field) is
consistent with the experimental information and a correlated electron
picture.

In the superconducting regime
$x \approx$ 0.3, the emerging picture is quite the opposite.  The specific heat
coefficient is ordinary, with the reported values\cite{lorenz,euland}
clustering around $\gamma$= 12-13 mJ/mol-K$^2$  
indistinguishable the band value $\gamma_{\circ}$
$\approx$ 13 mJ/mol-K$^2$.  In addition, there is 
no local moment (Curie-Weiss) contribution to the susceptibility,
and other evidence of enhanced properties is lacking (magnetic field 
dependence of the resistivity is small, for example).  In short,
evidence of substantial correlation effects due to the anticipated strong
on-site Coulomb repulsion $U$ is difficult to find for $x <$0.5.  
Moreover, as discussed in the Introduction, the $x$=0
endpoint is not a Mott insulator, but rather a nonmagnetic metal.\cite{private}

It is essential to begin to
reconcile the microscopic model with observations.
There are several indications that the simple single band picture is
oversimplified, one of the most prominent being that there is no evidence
that the $a_g$ state is significantly different in energy from the 
$e_g^{\prime}$ states, {\it i.e.} the $t_{2g}$ degeneracy is still essentially
in place.  Due to the form of dispersion in the CoO$_2$ layer, it is still
the case that holes doped into the band insulator NaCoO$_2$ go into the
$a_g$ band, making it viable to use a single band model in the small $1-x$
regime with a rather robust value of $U$, with a value of $U \approx $ 3  eV
possibly being sufficient to account for correlated electron behavior
($W \approx 1.5$ eV).  

The $x <$ 0.5 regime seems to require reanalysis.  It is quite 
plausible, based both on the LDA band structure and the observed properties,
that for $x <$ 0.5, the system crosses over into a three-band regime where
the full $t_{2g}$ complex of states comes into play.
The multiband nature tends to mitigate correlated behavior in at least two
ways.  Firstly, doped carriers that go into a multiband complex may
simply find a smaller ``phase space'' for approaching or entering the
Mott-Hubbard insulating phase, as the carriers have more
degrees of freedom.  Not completely separate, perhaps, is the extensive
study of Gunnarsson, Koch, and Martin,\cite{OG1,OG2,OG3}
that strongly suggests that
in a multiband system of N bands, the effective repulsion strength
becomes $U^{eff} = U/\sqrt{N}$.  For these cobaltates with carriers in
the $t_{2g}$ bands, $N$=3, and $U_c \leq$ 3 eV would become $U_c^{eff}
\rightarrow$ 3 eV/$\sqrt{3} \sim W$, and correlation effects diminish
considerably.  Secondly, screening will increase as hole doping occurs
from the band insulator $x$=1, reducing -- perhaps strongly -- the 
intra-atomic repulsion $U$ to a value near $W$..

\section{Summary}
In this paper we have begun an analysis, coupled with 
close attention to the observed
behavior, of the strength of correlation effects in this
cobaltate system that superconducts when hydrated.  In this initial work,
we have used the mean field LDA+U method to evaluate the 
effects of Hubbard-like interactions in NxCO, and find charge
disproportionation and a Mott insulating state for Coulomb repulsion
$U$ = 3 eV or less, for both $x=1/3$ and $x=2/3$, when
fluctuations are neglected.  Ferromagnetic tendencies for small $U$
evolve to nearest neighbor antiferromagnetic coupling $J \approx 11$ meV
for $U \sim$ 5 eV, at least if charge disproportionation occurs.
The only insulating phase reported so far has been at $x$=1/2, with strong
evidence\cite{cavaCO} that it is due to
charge disproportionation and charge order (and probably magnetic order).

The $x=\frac{1}{3}$ LDA
FS has been described, following the presumption (based on the
observation of at most moderately correlated behavior) that 
2D fluctuations will restore the paramagnetic metallic state.  There are
strong indications however that strong interactions, clearly evident
for $x >$ 1/2, have become muted in the regime where superconductivity
appears.
On the one hand, the electronic structure and FS indicate 
that multiband effects must be considered in this regime, which in itself 
will decrease the effective repulsion $U$.  Independently, $U$ will be
decreased by screening as the system becomes increasingly metallic.
On the experimental side, the behavior of both the magnetic susceptibility
and the linear specific heat coefficient point to a lack of ``enhanced''
behavior for $x \approx$ 0.3.

\section{Acknowledgments}
We have benefited from many stimulating discussions 
with R. T. Scalettar and
R. R. P. Singh, and clarifying communications with H. Alloul,
R. Cava, B. C. Sales, D. Mandrus, K. Takada, and J. M. Tarascon.  
J. K. was supported by National Science Foundation Grant 
DMR-0114818.  K.-W. L. was supported by DE-FG03-01ER45876.
W. E. P. acknowledges support of the Department of Energy's
Stewardship Science Academic Alliances Program.


\end{document}